# Networks of Innovation in 3D Printing

Harris Kyriakou, Steven Englehardt, and Jeffrey V. Nickerson

Stevens Institute of Technology
Hoboken, NJ
{ckyriako, sengleha, jnickerson}@stevens.edu

Innovation inside companies is difficult to see. But an emerging online community of inventors who publicly post 3D CAD drawings of their work provide a way to observe – and perhaps amplify – innovation. In this paper we analyze the network structure of Thingiverse, a website oriented toward 3D printing. This form of printing blurs the line between creating information and manufacturing objects: drawings can be sent to devices that build 3D objects out of many materials, including resin, ceramics, and metal [1].

Analysis of patent networks has suggested that the betweenness centrality of a patent contributes to its later success [2]. It is not only the structure of the network that determines success, but also what is referred to as *technological distance* [3]. Analysis of patents, though, has an important limitation: the patent references likely represent old inventions, and the references may be retrospective, the result of examiners' suggestions, rather than chronicles of real influence. In the case of Thingiverse, participants link their designs to parent designs. Essentially, they remix other CAD drawings, and their links are very direct acknowledgement of the information they have used (for more on remixing, see [4]). Many times inheritance is single, from one other design, but sometimes the participants inherit from more than one design. These, then, are combinations. From patent literature, and from other theoretical analyses [5], as well as our own past research [6] we expect that these combinations are worthy of closer examination.

As an exploratory study, we analyzed the structure of Thingiverse links. Specifically, we looked first for derived designs with more than one parent: as of June 2012, 281 out of 19,700 total designs had more than one parent. We then analyzed these designs along two dimensions. The first was normalized betweenness centrality. The second was a measure of conceptual independence. We wanted to know how dissimilar the parents were, under the expectation combinations of similar designs would be less novel than combinations of dissimilar designs.

Designers labeled their creations with pertinent tags. In order to calculate the independence score of each design, we used the ratio of the intersection of all parent tags of a design divided by the union of the parent tags. If a design was derived from very similarly tagged designs, its *Independence Score* would be close to zero, whereas radical designs combining ideas from different segments would score close to one (this is a Jaccard distance calculation, generalized to measure the collective independence of many parents):

$$Independence\ Score = 1 - \frac{|P_1 \cap P_2 \cap ... \cap P_n|}{|P_1 \cup P_2 \cup ... \cup P_n|}$$

Figure 1 shows the designs plotted along two dimensions. We expect the upper quadrant will contain the more novel designs: as an exploratory study we picked five designs at random from each quadrant and analyzed them. Our observations suggest the more interesting designs occur in quadrant three, and in particular the quadrant three designs with higher independence scores. Figure 2A shows a low left quadrant prosaic design with low independence and betweenness scores. By contrast, Figure 2B shows from the high right quadrant the merge of a grenade container derived from a novelty lemon cap and a parametric screwable box – obviously dissimilar sources of inspiration.

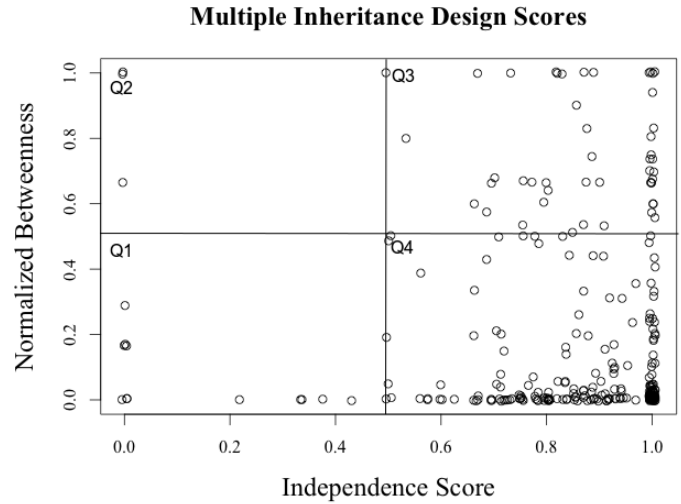

**Figure 1. Multiple Inheritance Design Scores**

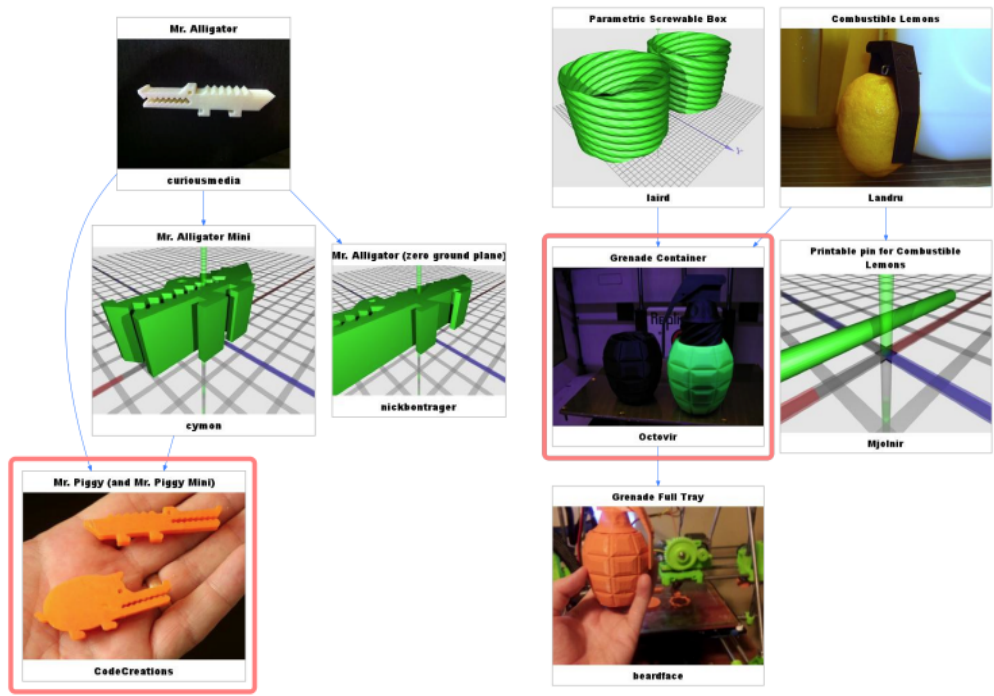

**Figure 2A (left) and 2B (right). These are examples of design trees in the lower left (Q1) and upper right (Q2) quadrants of Figure 1.**

Also in our randomly selected set was a robot derived from two designs: one based on a custom circuit board not widely available, and another that is based on the popular Arduino board, as shown in Figure 3.

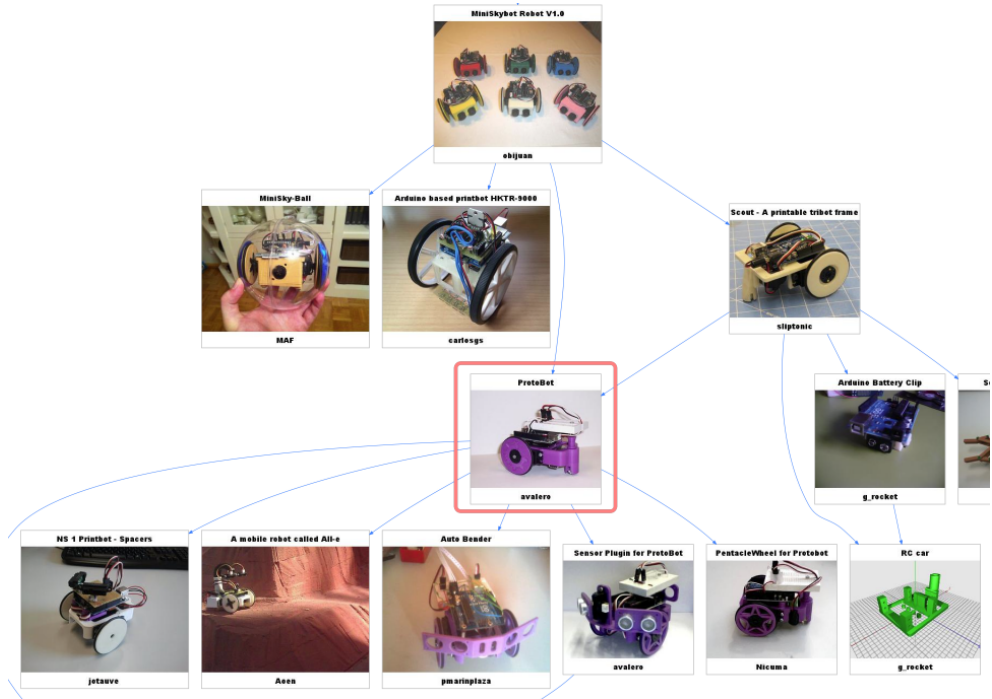

**Figure 3. Protobot Parent-Child Tree**

These exploratory results suggest that analysis of remix network structure may provide ways of tracing innovation processes and detecting the emergence of new ideas, combination of disparate ideas. Such understanding is a goal in its own right. As part of this understanding, we might refine measures of design inbetweenness and measures of conceptual independence. In addition, it may be possible to encourage innovation by guiding interested inventors toward potential pairs of structurally separated and conceptually diverse ideas. The transparent system of innovation provided by Thingiverse may allow us to amplify great ideas by prompting new possibilities.

While innovation networks abound, few have the fast pace of the Thingiverse network: ideas are inherited from others, and then manufactured, affecting the world soon after they are published. The creative combination of designs we found suggests the value in an open approach to innovation in which copyright and patents are abandoned for fast, breakneck innovation.

**Acknowledgments**

This work was supported under National Science Foundation award IIS-0855995 and award IIS-0968561.